\documentclass[prd, aps, superscriptaddress, preprintnumbers, twocolumn, floatfix, nofootinbib]{revtex4}

\usepackage{amsfonts}
\usepackage{amsmath}
\usepackage{amssymb}
\usepackage{bm}
\usepackage{dcolumn}
\usepackage{epsfig}
\usepackage{graphicx}   
\usepackage{graphics}
\usepackage[latin1]{inputenc}
\usepackage{latexsym}
\usepackage{rotating}
\usepackage{hyperref}

\usepackage{amsfonts}
\usepackage{amsmath}
\usepackage{amssymb}
\usepackage{bm}
\usepackage{dcolumn}
\usepackage{epsfig}
\usepackage{graphicx}
\usepackage{graphics}
\usepackage[latin1]{inputenc}
\usepackage{latexsym}
\usepackage{rotating}
\usepackage{hyperref}

\newcommand\be{\begin{equation}}
\newcommand\ba{\begin{eqnarray}}
\newcommand\ee{\end{equation}}
\newcommand\ea{\end{eqnarray}}

\begin{document}

\title {Initial Conditions for Inflation - A Short Review}

\author{Robert Brandenberger}
\email{rhb@physics.mcgill.ca}
\affiliation{Physics Department, McGill University, Montreal, QC, H3A 2T8, Canada, and \\Institute for Theoretical Studies,
ETH Z\"urich, CH-8092 Z\"urich, Switzerland}

\date{\today}

\begin{abstract}

I give a brief review of the status of research on the nature
of initial conditions required to obtain a period of cosmological inflation.
It is shown that there is good evidence that in the case of large field
models, the inflationary slow-roll trajectory is a local attractor in
initial condition space, whereas it is not in the case of small field
models.

\end{abstract}

\maketitle

\section{Introduction} 

The inflationary scenario \cite{Inflation} has become the current paradigm of early
universe cosmology. A period of exponential expansion of space solves a number
of fine-tuning problems of Standard Big Bang cosmology, in particular the horizon
and flatness problems. Without inflation, finely tuned initial conditions are required
to explain the observed degree of spatial flatness, and acausal initial condition correlations
are required in order to explain the observed near isotropy of the cosmic microwave
background (CMB).

In order for these achievements of cosmological inflation to be counted as real
successes, inflation should arise without having to impose very special initial
conditions. At the present time, there are rather conflicting statements on the
naturalness of initial conditions for inflation, from articles which claim that there
is a very serious problem \cite{Ijjas} to those which claim that inflation arises
very naturally \cite{Nomura, Linde}.

The purpose of this article is to review the status of the initial condition issue for
cosmological inflation, and to explain why different authors obtain such diverging
opinions on the issue. To be specific, I will focus on inflation obtained from matter scalar
fields minimally coupled to General Relativity as the theory of space-time. 
The outline of this review is as follows. We first discuss early
work which studied the initial condition issue in the context of homogeneous
and isotropic cosmologies. The real issue, however, is whether inflationary
expansion can occur if inhomogeneous initial conditions for both matter fields
and the metric are allowed. This will be discussed in the third section of the
paper. We conclude with a discussion section.
 
\section{Initial Conditions in Mini-Superspace}

In this review I will focus on inflation obtained by means of scalar matter
fields minimally coupled to Einstein gravity. A period of inflation
corresponds to a period of accelerated expansion of space. In the
context of Einstein gravity, matter with an equation of state $p < \frac{1}{3} \rho$
is required in order to obtain such expansion. 

The inflationary scenario provides a theory for the origin of
structure in the universe which can explain CMB anisotropies and
the distribution of galaxies on large scales \cite{Mukh} (see also
\cite{Sato, Press}). Comparison between the predictions of this theory
and the latest CMB observations \cite{Planck} tells us that the
expansion rate of space was nearly exponential during inflation.

Since exponential expansion corresponds to a de Sitter phase of
expansion, the initial conditions question for inflation was first
discussed in the context of de Sitter space. De Sitter space can
be obtained by adding a cosmological constant $\Lambda$ to
the gravitational action. At a classical level, one can then prove
``no-hair" theorems for de Sitter space which show that
classical fluctuations about de Sitter space redshift and space-time
approaches pure de Sitter \cite{Hawking}. Assuming matter which
satisfies the ``strong energy condition", the no-hair theorem was
proved in \cite{Wald} for an isotropic background metric, and by
Starobinsky \cite{Starob} (see also \cite{no-hair-anis}) for
anisotropic backgrounds.Extensions to the case of matter
not satisfying the strong energy condition were recently
studied in \cite{Shahin}. The stability of de Sitter space-time
to inhomogeneities with vanishing Weyl curvature tensor
\cite{Weyl}, to the addition of linear gravitational waves
\cite{Starob2} and to linear cosmological fluctuations \cite{scalar}
was also studied. The no-hair conjecture has also been
proved at nonlinear level for a model containing dust matter
plus a bare cosmological constant \cite{Tavakol}. Further extensions and general
considerations were made e.g. in \cite{ext}. All of these
analyses assume that matter obeys the strong energy condition.
If this assumption is dropped, then there is the possibility
of unstable modes \cite{Barrow}. 

Quantum mechanically, the question of stability of de Sitter
space-time is open. It is possible that infrared instabilities can
destabilize de Sitter (see e.g. \cite{Polyakov}), although this
issue is hotly debated (for an opposing point of view see e.g.
\cite{Marolf}). Studies of tensor fluctuations \cite{Woodard}
also indicate the de Sitter space-time is unstable, and the
onset of an instability due to scalar entropic fluctuations can
also be shown \cite{RHB}, although these studies are only
perturbative and hence cannot truly differentiate between
an instability of de Sitter and a finite renormalization of the
cosmological constant due to semi-classical effects.

However, the inflationary scenario is not the dynamics
of space-time in the presence of a bare cosmological
constant, but it is the cosmology of a scalar field with
a non-trivial potential energy term in the absence of a
cosmological constant (any cosmological constant which
might explain the current dark energy has negligible effects
in the very early universe). Hence, the studies of de Sitter
cosmic no-hair conjectures are not  directly applicable
to inflationary cosmology.

As already mentioned, we will focus on scalar field
realizations of inflation. For simplicity I will also assume
that the scalar field $\varphi$ has canonical kinetic term.
Thus, the action $S$ is given by
\be
S \, = \, \int d^4 x \sqrt{-g} \frac{1}{- 16 \pi G} R 
+ \frac{1}{2} \partial_{\mu} \varphi \partial^{\mu} \varphi - V(\varphi) \, ,
\ee
where $g$ is the determinant of the metric, $R$ is the
Ricci scalar, $G$ is Newton's gravitational constant, and $V$
is the potential energy function of $\varphi$. Space-time
indices are labelled with Greek letters. Accelerated expansion
of space is obtained when the potential energy of
$\varphi$ dominated over the kinetic energy ${\dot \varphi}^2$
and the tension energy $(\nabla \varphi)^2$. In order
for inflation to solve the problems of Standard Big Bang
cosmology which it was designed to solve (homogeneity
of the CMB, observed spatial flatness), the period of
inflation must be longer than about $50$ e-foldings.

In the following it will be important to distinguish between
{\it small field} and {\it large field} models of inflation. For
scalar field which have canonical kinetic term, {\it small field}
inflation corresponds to situations in which the scalar field moves
over a field value range $\Delta \varphi < d m_{pl}$, where $m_{pl}$
is the Planck mass and $d$ is a constant of order $1$ whose
precise value depends on the form of the potential $V(\varphi)$.
We speak of {\it large field} inflation if $\Delta \varphi > d m_{pl}$.

The initial slow-roll models of inflation \cite{NewInflation} 
(``New Inflation") were based
on the idea that $V(\varphi)$ is a symmetry breaking potential
of similar shape as the potential of the Higgs field, that thermal
effects initally localize $\varphi$ at the local maximum of
the potential located at $\varphi = 0$, and that $\varphi$ starts
to roll slowly towards the absolute minimum of the potential
located at $|\varphi| = \eta \ll m_{pl}$. A first problem arises
since a simple double well potential 
\be
V(\varphi) \, = \, \frac{1}{4} \lambda \bigl( \varphi^2 - \eta^2 \bigr)^2
\ee
does not admit self-consistent slow roll solutions. Slow-roll
solutions can be obtained if the potential near $\varphi = 0$
is smoothed out by taking it to be of the Coleman-Weinberg \cite{CW}
form. However, it was soon
realized \cite{chaotic, MWU} that in order for
the cosmological fluctuations produced during inflation not to be
too large in amplitude, the coupling constant in the potential
$V(\varphi)$ as well as those coupling $\varphi$ to other matter
fields must be so small that $\varphi$ will not be in thermal
equilibrium at early times. Hence, the justification for starting
the evolution at $\varphi \sim 0$ falls away. 

The initial condition problem for small field models of inflation
was studied in detail in \cite{Piran1} at the level of a classical
phase space analysis. It was shown that in order to obtain
a sufficiently long period of inflationary expansion, the initial
scalar field velocity has to be finely tuned (i.e. tuned to be
much smaller than the characteristic value
$|{\dot \varphi}| = H^2$) even if the initial
value of the field itself is set by hand to be very close to zero.

Whereas this fine tuning of the initial scalar field velocity seems
unnatural for simple potentials, there is one context in which
it is natural, namely if the initial conditions for the small field
inflationary phase are set by tunneling from a false vacuum
\cite{tunneling}. This scenario cannot be realized with
renormalizable scalar field potentials, but it may well arise
in effective field theories obtained from string theory. What
is required is a potential for which $\varphi = 0$ is a metastable
minimum, and which has a sufficient degree of flatness for
field values $|\varphi|$ beyond the nucleation value to obtain
enough slow-roll inflation.

Another context in which the initial scalar field velocity might
vanish is if the initial conditions for Minkowski space-time
evolution are set via an analytic continuation from a Euclidean
quantum gravity region. This is the quantum cosmology
approach to the initial condition problem of small field
inflation. The problem is that different ans\"atze for the
wave function of the universe (specifically the Hartle-Hawking \cite{HH}
wavefunction or the ``tunneling" wavefunctions \cite{LiVi})
give very different results for the probability of inflation when applied
to the same Lagrangian system (see e.g. \cite{HHH} for
a computation of the probability given the wave function of \cite{HH}).

The situation is very different in large field models of inflation. As
early work in \cite{AB1} already hinted at, and as was studied
in mode detail in \cite{Kung}, the inflationary slow-roll
trajectory is a local attractor (see \cite{Carroll1} for a more mathematical
discussion of the meaning of the term ``attractor'' in this context)
in initial condition space, at least at
the level of homogeneous and isotropic cosmology
(see also \cite{Mukh3}). In large field
inflation models there is enough time for the excess kinetic energy
compared to the energy required during slow-rolling has enough
time to redshift, whereas this is not the case for small field inflation.

The difference in the dynamics between small field and large field
inflation can be illustrated with phase space diagrams, as
shown in Figs. 1 - 4. In all four figures, the horizontal axis represents
the value of the scalar field. Fig. 1 is a sketch of the potential
energy density function (vertical axis)  assumed in small field models,
Fig. 3 is the analog for large field models of inflation.
Figs. 2 and 4 are sketches of the phase space dynamics for a 
small field inflation model (Fig. 2) and a large field model (Fig. 4). The
vertical axes represent the field momentum. 
The arrows on the trajectories indicate
the evolution in time. It is clear from Figs. 2 and 4 that the slow-roll
trajectory labelled with ``SR" is a local attractor in initial condition
space in the case of large field inflation, but not in the case of small
field models. The difference in the likelihood of inflation (in the context
of homogeneous and isotropic cosmology) between large and small
field models of inflation has also been more recently confirmed in
\cite{Carroll2}.

\begin{figure*}[t]
\begin{center}
\includegraphics[scale=0.4]{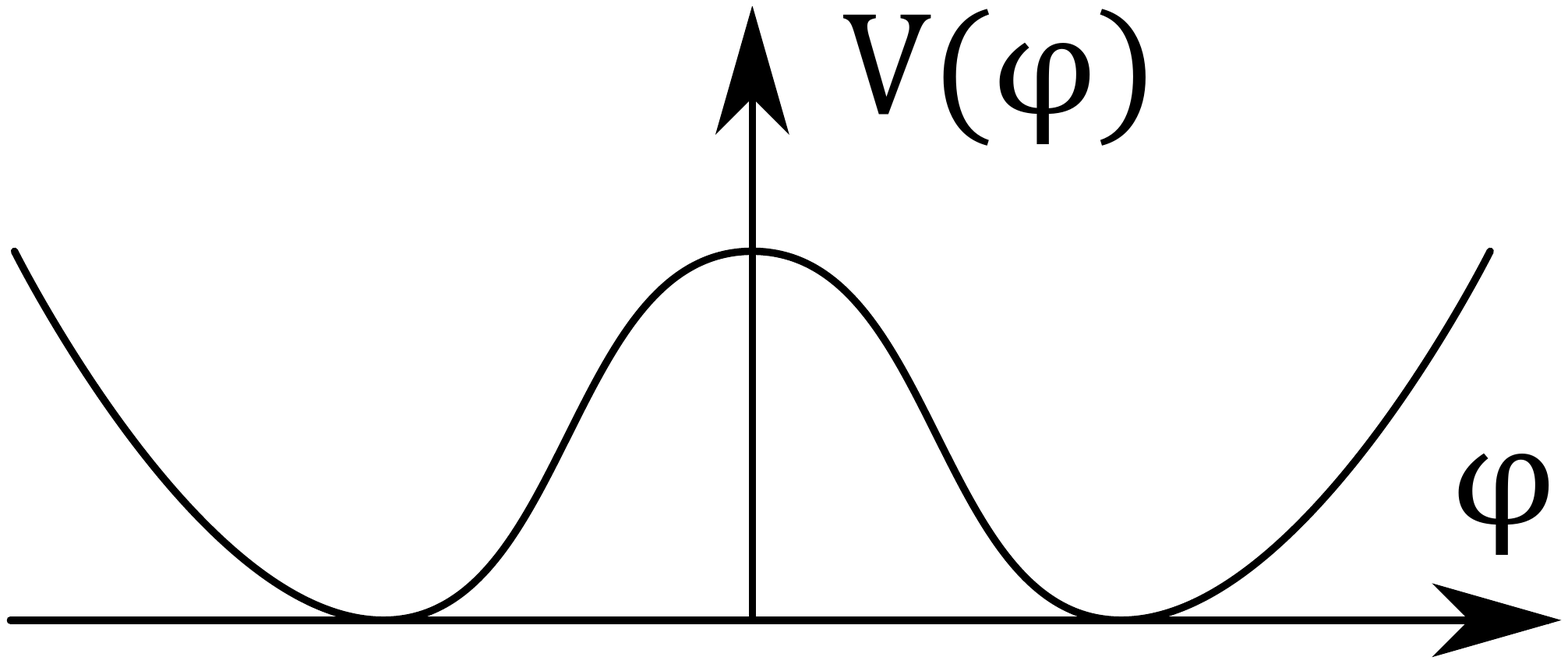}
\caption{Sketch of the potential energy function (vertical axis) as a function of the
scalar field value (horizontal axis) for small field inflation models. The initial
conditions for $\varphi$ are taken to be close to the top of the unstable
state at $\varphi = 0$.}
\end{center}
\end{figure*}

\begin{figure*}[t]
\begin{center}
\includegraphics[scale=0.4]{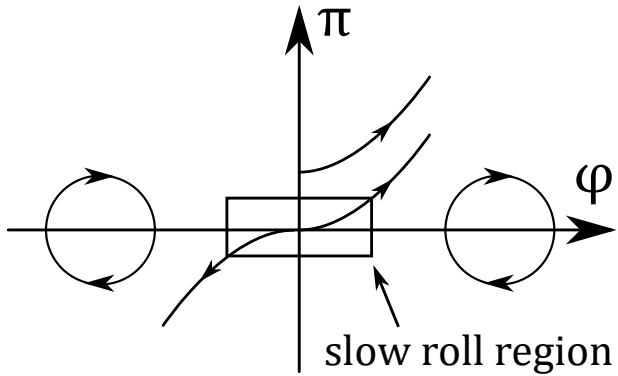}
\caption{Sketch of phase space trajectories in small field inflation models.  The horizontal axis is the field,
the vertical axis is the momentum (equivalently the time derivative of the field). Typical
initial conditions with nonvanishing $\pi$ do not lead to inflation.}
\end{center}
\end{figure*}

\begin{figure*}[t]
\begin{center}
\includegraphics[scale=0.4]{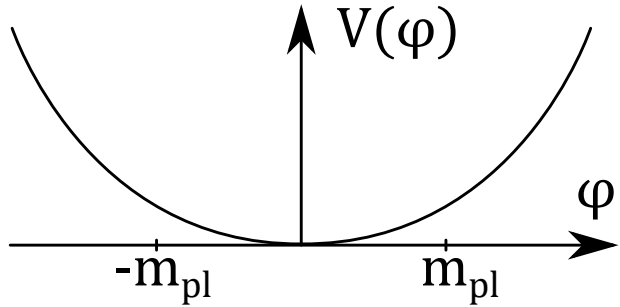}
\caption{Sketch of the potential energy function (vertical axis) as a function of the
scalar field value (horizontal axis) for large field inflation models.}
\end{center}
\end{figure*}

\begin{figure*}[t]
\begin{center}
\includegraphics[scale=0.4]{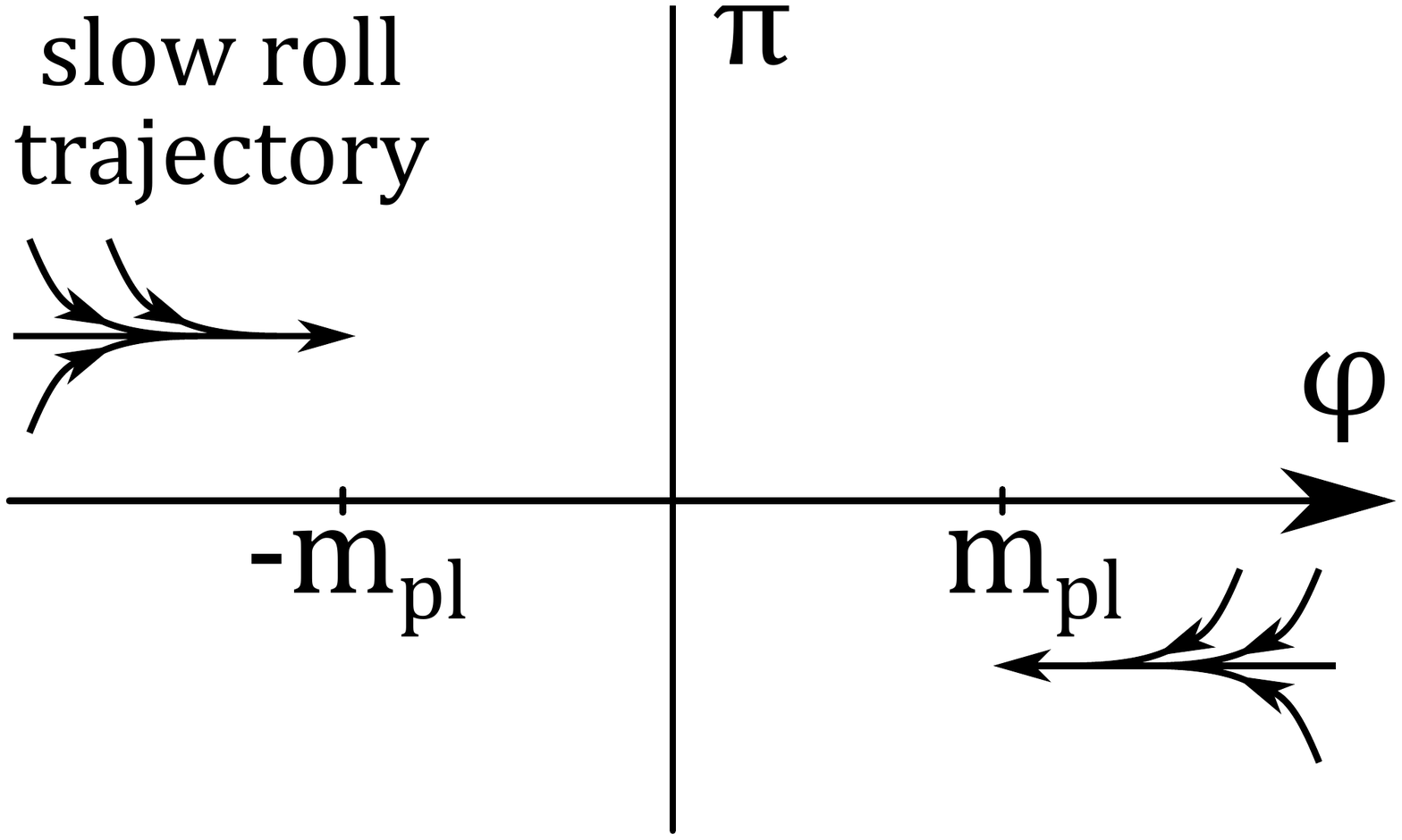}
\caption{Sketch of phase space trajectories in large field inflation models. 
The axes are as in Figure 2. The slow-roll trajectories (the
horizontal trajectories at $|\varphi| > m_{pl}$ are now
local attractors in initial condition space.}
\end{center}
\end{figure*}

The above considerations have been at the level of classical dynamical
systems. Implicit in the analysis is that all initial conditions which have
a fixed energy density are equally likely, and as initial density we take
a density when the classical dynamical desription becomes justified (e.g.
the Planck density). However, in a quantum universe the wave function
of the universe may not give equal probability to different initial conditions
with the same energy density, but may prefer special initial conditions.
On these grounds it was argued a long time ago that the initial conditions
for inflation (at the mini superspace level) are very likely \cite{Hawking2}.
On the other hand, Penrose \cite{Penrose} and others \cite{GT} have
argued that inflation is extremely unlikely. The different results are due
to different measures chosen. Specifically, the analysis of \cite{Penrose, GT}
uses a measure on phase space configurations today, and asks what set
of configurations similar to the ones we observe now have come from 
inflationary initial conditions, whereas in the analysis of \cite{Hawking2}
one is studying measures on the set of initial conditions, a procedure which
is more appropriate if one has in mind standard Cauchy evolution. In
fact, it has been shown \cite{Corichi} (see also \cite{Corichi2}) that the
measure gets squeezed by time evolution into regions of phase space
which yield inflation. However,
general difficulties in making arguments based on a measure of initial
configurations were discussed in \cite{Wald2}.

\section{Initial Conditions for Inhomogeneous Dynamics}

A discussion of the probability to obtain inflation based on a mini-superspace
analysis is only a very limited aspect of the initial conditions problem
for inflation. If we assume homogeneity from the outset, then there
is no horizon problem and inflation is not needed to explain the near
isotropy of the CMB (it is still needed to explain the spatial flatness
and to provide causal initial conditions for fluctuations). Thus, we must
now consider how likely it is to obtain inflation from general inhomogeneous
initial conditions. This issue is not yet completely settled. There are
authors who argue that exponential fine tuning of initial conditions
are required \cite{Trodden1, Trodden2}, whereas others argue that
even in the presence of matter and metric fluctuations the slow-roll
trajectory of large field inflation is a local attractor in initial condition
space \cite{ABM, Matzner1, Muller, Kung, Feldman}. In the following
we will only discuss large field inflation models.

Why are such different conclusions reached? The analysis of
\cite{Trodden1, Trodden2} starts from the assumption that
``homogeneity" over a length scale of more than $H^{-1}$, where
$H$ is the Hubble expansion rate at the beginning of inflation,
is required in order to obtain inflation. By ``homogeneity" it is
here meant that the energy density has to be dominated by
the almost constant mode of the scalar field. It is then argued
that it is very unlikely to have such initial conditions. This
argument, however, clearly will give only a lower bound on
the probability of inflation. Since inhomogeneities redshift,
but a homogeneous field only slowly rolls and hence decays
much more slowly as a function of time than the fluctuating
modes, Hence, homogeneity over a few Hubble patches
is a sufficient condition to obtain inflation, but clearly it is
not a necessary one.

In fact, it is argued in \cite{ABM, Muller, Kung, Feldman}
that even if the initial energy density was dominated by inhomogeneous
modes (inhomogeneity scale smaller than $H^{-1}$) the universe
is likely to eventually enter a period of inflation provided that there
is some power in the quasi-homogeneous mode (homogeneous
on scale $H^{-1}$) \footnote{Note that having power on super-Hubble
scales is expected. In fact, the absence of such fluctuations would
require acausal initial conditions.} The argument is as follows. Whatever the
initial conditions are, the universe will expand. During this
expansion phase the energy in the inhomogeneous modes will
redshift whereas the zero mode of $\varphi$ will only slowly
roll. Hence, as long as there is sufficient power in the zero
mode, it will eventually come to dominate and the universe will
start to inflate. Thus, the inflationary slow-roll trajectory is
a local attractor in initial condition space, even in the presence
of inhomogeneities. This argument, however, misses
potential back-reaction effects. To second order in the
amplitude of the fluctuating modes, these modes have
an effect on the background. In particular, it is possible
that they will effectively destroy the background and hence
prevent the onset of inflation.

It has in fact been shown numerically \cite{ABM} that in the
absence of metric fluctuations the slow-roll trajectory for
large field inflation is a local attractor. The attractor basin is
in fact very large - the initial density in the inhomogeneous
fluctuations may be orders of magnitude larger than that in
the zero mode. A much improved numerical analysis of
this problem has recently been published in \cite{Easther}.

However, it is not consistent to neglect metric fluctuations.
One may worry that the presence of initial metric fluctuations
may lead space-time to collapse into a gas of black holes rather
than lead to an expanding cosmology. On the other hand,
the work of \cite{Feldman} showed that even in the presence
of metric fluctuations (linearized joint fluctuations of metric
and matter) the large field slow-roll trajectory remains an
attractor in initial condition space. 

To go beyond linear theory, methods of numerical relativity
are required. The initial numerical studies of the onset of
inflation with inhomogeneous initial conditions were performed
in \cite{Piran2, Piran1} in the case of small field inflation models,
and in \cite{Matzner1} in the case of large field inflation. Both
works used one space-dimensional codes. It was shown that
inflation was very unlikely in the case of small field inflation,
but that it is an attractor in initial condition space for large field
inflation. 

With the improvements in numerical relativity codes, and
the availability of more computational power, it has become
possible to perform simulations in three spatial dimensions.
Early work in three space dimensions is due to \cite{Matzner2},
and a more recent study was recently published in \cite{East}. These
studies all come to the conclusion that the slow-roll trajectory
for large field inflation is a local attractor given the set of initial
conditions which were studied. Specifically, in the recent work
of \cite{East} it was found that a Hubble patch with initial conditions in
which the inhomogeneities (modelled as a set of Fourier
modes with sub-Hubble wavelength) have an energy
density which exceeds that of the background by a
factor of $10^3$ will eventually undergo inflation, as
long as the average spatial curvature is not postive,
and as long as the field range remains in the slow-roll
regime. If the field values entered the non-slow-roll region,
it was found that sometimes the patch will not inflate.  Thus,
whereas the inflationary slow-roll trajectory is a local attractor,
it is not a global attractor. Another new code is in
development \cite{Lim} and will be used to study the onset
of inflation.

\section{Conclusions}

As this review has shown, there has been a lot of work concerning
initial conditions for inflation. The author feels that there is now a
body of analytical and numerical work supporting the claim that
in the case of large field models
the slow-roll trajectory is a local attractor in initial condition space.

It is important to distinguish the issue of initial conditions for inflation
from that of the initial conditions for the cosmological fluctuations
and gravitational waves in an inflationary background. Concerning
the latter, one usually argues that the Bunch-Davies \cite{BD}
vacuum state is an attractor in the space of states (as far as
the evolution of correlation functions is concerned) (see e.g.
\cite{Hill} for some early work and \cite{YiWang} for a recent
review). The question of initial conditions for fluctuations is
related to the ``trans-Planckian problem'' for cosmological
fluctuations which will be touched on below.

Some researchers discount the entire discussion of initial conditions
for inflation by saying that as long as there will be one patch which
starts to inflate, its physical volume will rapdily come to dominate
the Universe.  In particular, if the field is allowed to explore the
region in which stochastic forces driving the field up the potential
becomes larger in magnitude than the classical force driving
the field down the potential \cite{Starob}, then the physical volume of space
will be dominated by inflationary patches. However, what ``dominate'' means 
depends on the measure one considers, and it is claimed in \cite{Trodden2}
that the probability of inflation remains low even if one takes
the exponential growth of inflating regions into account. This debate
may be academic, however, since small modifications of the
potential at super-Planckian field values can easily eliminate the
region of stochastic inflation \cite{Mukh3}. As a bottom line, it is reassuring that
a local dynamical systems analysis seems to show that the initial conditions
for inflation do not have to be finely tuned.

 Even if the inflationary slow-roll trajectory is a local attractor in
 initial condition space, it is not a global attractor. If the inhomogeneities
 are too large, than we would expect space to collapse into a gas
 of black holes, leading to a picture of the early universe discussed
 in \cite{Banks} or \cite{Veneziano}. The author feels, however, that
 the inflationary slow-roll trajectory being a local attractor is already
 a  promising achievement.
 
 Having concluded that the slow-roll trajectory in large field inflation
 is a local attractor in initial condition space, one must keep in mind
 that there are problems in embedding large field inflation into a
 ultra-violet finite quantum theory such as string theory. Axion
 monodromy inflation \cite{Eva} appears to be the most promising
 route, but there are concerns whether such models are safe
 against back-reaction effects \cite{Hebeker} and considerations
 based on the ``Weak Gravity Conjecture'', considerations
 which rule out large field inflation in other axion inflation models 
 \cite{Rudelius}. If these problems hold up and one is forced into small field
 inflation, then the initial condition problem for inflation would become
 rather severe.
 
 It is important to keep in mind that the inflationary scenario is at
 best an incomplete picture of the very early universe. Inflation, at
 least in the context of scalar field matter coupled to Einstein gravity,
 is known to be past incomplete \cite{Borde}. This implies that we
 need to go beyond inflationary cosmology if we really want to 
 understand the very earliest moments of the universe. 
 
 There are other conceptual problem of inflation (see e.g. \cite{RHBrev}).
 For one, if inflation last more than $70$ e-folding times, then the
 wavelengths of all scales which are being obsered today are
 smaller than the Planck length at the beginning of the period of
 inflation. As shown in \cite{Jerome} and many followup papers,
 this gives rise to the ``trans-Planckian'' problem for cosmological
 fluctuations: effects of physics in the ``zone of ignorance'' of
 length scales smaller than the Planck length may well effect the
 nature of the observed fluctuations. Another issue concerns the
 sensitivity of the inflationary mechanism - almost constant potential
 energy density driving accelerated expansion, to our ignorance
 of what renders quantum vacuum energy gravitationally insert.
 
 These and other conceptual issues have led to the development
 of alternative early universe scenarios such as the Pre-Big-Bang
 \cite{PBB}, the Ekpyrotic scenario \cite{Ekp}, the ``Matter Bounce''
 \cite{Fabio} and String Gas Cosmology \cite{SGC}, to mention only
 a few. Whereas these scenarios do not suffer from the
 trans-Planckian problem for fluctuations, they all have their own
 conceptual problems. In particular, the initial conditions required for several 
 of them (e.g. the Pre-Big-Bang and the Matter Bounce scenarios)
are not attractors in initial condition space \cite{PP} (on the other hand, the
trajectory in the original Ekpyrotic scenario is an attractor \cite{Gratton}).

\section*{Acknowledgement}
\noindent

The author is grateful to Yi Wang for producing the figures, and to
L. Berezhiani, S. Carroll, A. Corichi, V. Mukhanov, T. Rudelius, M. Sheikh-Jabbari 
and T. Vachaspati for useful discussions and feedback.
He wishes to thank the Institute for Theoretical Studies of the ETH
Z\"urich for kind hospitality. RB acknowledges financial support from Dr. Max
R\"ossler, the Walter Haefner Foundation and the ETH Zurich Foundation, and
from a Simons Foundation fellowship. The research is also supported in
part by funds from NSERC and the Canada Research Chair program.

\end{document}